\documentclass[prl,twocolumn,showpacs]{revtex4}

\usepackage{graphicx}

\begin{document}                          

\title{Magnetoresistance of atomic-sized contacts: 
an ab-initio study} 

\author{
A.~Bagrets$^{1,2}$, N.~Papanikolaou$^{3}$, and I.~Mertig$^{2}$ }

\affiliation{
$^1$Max-Planck-Institut f\"ur Mikrostrukturphysik, D-06120 Halle, Germany\\
$^2$Martin-Luther-Universit\"at Halle, Fachbereich Physik, D-06099 Halle, Germany \\
$^3$Institute of Microelectronics, NCSR "Demokritos", GR-15310 Athens, Greece }
 
\date{\today}

\begin{abstract}
The magnetoresistance (MR) effect in metallic atomic-sized contacts
is studied theoretically by means of first-principle electronic 
structure calculations. We consider three-atom chains formed from Co, Cu, Si, and Al atoms 
suspended between semi-infinite Co leads. We employ the screened Korringa-Kohn-Rostoker 
Green's function method for the electronic structure calculation
and evaluate the conductance in the ballistic limit using the Landauer approach.
The conductance through the constrictions reflects 
the spin-splitting of the Co bands and causes high MR ratios, 
up to 50\%. The influence of the structural changes 
on the conductance is studied 
by considering different geometrical arrangements of atoms forming the
chains. Our results show that the conductance through $s$-like states
is robust against geometrical changes, whereas the transmission
is strongly influenced by the atomic arrangement if $p$ or $d$ states 
contribute to the current.  
\end{abstract}


\maketitle                                                                                             

The investigation of electron transport through metallic atomic-sized
contacts has attracted a lot of attention during the last 15 years,
and most achievements are summarized in a recent review
~\cite{Ruitenbeek}.
Using up-to-date experimental techniques, 
such as mechanically controllable break junctions (MCBJ)~\cite{MCBJ}
or scanning tunneling microscopy~\cite{Rubio,Ohnishi}, 
it is possible to fabricate nanocontacts with 
a single atom, atomic chain or a molecule\cite{vanRuitenbeek_Nature}
in the constriction. The experiments reveal step-like changes 
of the conductance upon elongation 
of the nanocontacts~\cite{Ruitenbeek,Rubio,Ohnishi,MCBJ}. 
In case of noble (Au, Ag, Cu)~\cite{Brandbyge1,Krans,Ludoph} 
and alkali metals (Li, Na, K)~\cite{Krans,Ludoph,Yanson} 
conductance histograms show a dominant peak very close 
to one conductance quantum, $G_0 = 2e^2/h$, 
and smaller peaks close to integer conductance quanta.
However, in case of transition metals 
the situation differs significantly~\cite{Ludoph,Enomoto} 
and a broad distribution of conductance values is usually obtained.

To describe the nanocontacts theoretically, 
several methods have been developed by many 
groups during the last years.  An approach based on 
the tight-binding (TB) Hamiltonian adapted for a nanocontact geometry 
was proposed in Refs.~\onlinecite{Cuevas,TB-methods}.
An important conclusion of
TB models is that the conductance of single-atom  
contacts is related to the number of valence orbitals
of the contact atom available at the Fermi energy~\cite{Cuevas,Scheer}. 
Lang and coworkers~\cite{Lang,Lang_Al, Lang_Na, Lang_C}  
and Kobayashi and coworkers~\cite{Kobayashi_1,Kobayashi_2}  
studied the single-atom contacts~\cite{Lang}, 
atomic chains of Al~\cite{Lang_Al,Kobayashi_1}, 
Na~\cite{Lang_Na,Kobayashi_2}, and~C~\cite{Lang_C} 
using {\it ab-initio} calculations based on density functional (DF) theory 
with jellium electrodes. The formation mechanisms of atomic
chains made from different types of elements 
such as Ni, Pd, Pt, Cu, Ag and Au
were investigated by means of molecular dynamics simulations~\cite{Bahn}.   
These studies were triggered  by the experimental evidence of the formation 
of gold wires~\cite{Ohnishi}. Recently Mehrez {\it et al.}~\cite{Mehrez} 
and Brandbyge {\it et al.}~\cite{Brandbyge} presented fully self-consistent
DF calculations of the conductance of atomic-sized contacts treating
the electronic structure of both the contact and electrodes  
on the same footing.  

\begin{figure}[b]
\begin{center}
\includegraphics[scale = 0.18]{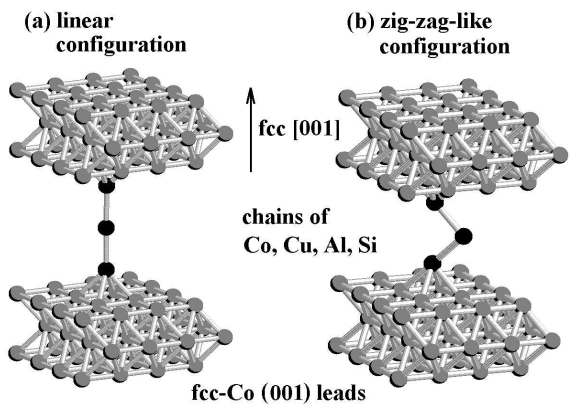}
\caption{
\label{fig1} Geometry of the nanocontacts: 
(a) linear configuration; (b) zigzag-like configuration.}
\end{center}
\end{figure}

 In this paper, we present {\it ab-initio}
calculations of the conductance of nanocontacts and focus on magnetic systems. 
We consider magnetic semi-infinite fcc (001) Co electrodes 
joined by nanocontacts of different materials. We assume that
they take the form of short three-atom Co-, Cu-, Si- or Al-chains 
suspended between the leads as shown in Fig.~\ref{fig1}.
Our aim is to investigate whether such hybrid systems could exhibit
a large magnetoresistance (MR) effect.
We wish to address the following theoretical questions: 
(i) What is the influence of transition metal electrodes 
on the conductance of nanocontacts? 
(ii) What is the effect of the geometrical
and electronic structure of the nanocontacts on the transport properties? 
(iii) What are favorable conditions to increase the MR 
through a constriction?

\begin{table*}
\caption{Magnetic moments ($\mu_B$) 
of atoms forming either linear or zigzag-like three-atom 
chains (see Fig.~\ref{fig1}) for parallel (P) and antiparallel (AP) 
orientation of the magnetic moments in the Co leads. 
The 'contact' atom of the chain sits above the (001) surface. 
The magnetic moment of bulk Co is $1.62~\mu_B$,
the moment of the Co (001) surface atom is $1.78~\mu_B$.}

\begin{ruledtabular}
\begin{tabular}{c|cccc|cccc|cccc|cccc}
{} &  \multicolumn{2}{c}{Co linear} &
\multicolumn{2}{c|}{Co zigzag} &
\multicolumn{2}{c}{Cu linear} &
\multicolumn{2}{c|}{Cu zigzag} &
\multicolumn{2}{c}{Al linear} &
\multicolumn{2}{c|}{Al zigzag} &
\multicolumn{2}{c}{Si linear} &
\multicolumn{2}{c}{Si zigzag} \\
{atoms} & P & AP & P & AP & P & AP & P & AP
& P & AP & P & AP & P & AP & P & AP \\
\hline
contact & $+$1.91 & $+$1.91 & $+$2.00 & $+$1.97 &
    $+$0.08 & $+$0.08 & $+$0.08 & $+$0.08 &  
    $-$0.09 & $-$0.09 & $-$0.10 & $-$0.09 & 
    $-$0.11 & $-$0.11 & $-$0.07 & $-$0.06 \\
central & $+$2.06 & $\phantom{+}$0.00 & $+$2.28 & $\phantom{+}$0.00 &
    $+$0.01 & $\phantom{+}$0.00 & $+$0.03 & $\phantom{+}$0.00 &  
    $\phantom{+}$0.00 & $\phantom{+}$0.00 
        & $\phantom{+}$0.00 & $\phantom{+}$0.00 & 
    $+$0.06 & $\phantom{+}$0.00 & $+$0.11 & $\phantom{+}$0.00 
\end{tabular}
\end{ruledtabular}
\label{table1}
\end{table*}

The structure of the nanocontacts studied below is presented in Fig.~\ref{fig1}. 
The experimental lattice constant $a_0 = 6.70$~a.u.\ of fcc Co was used 
in the calculations. In the first configuration (Fig.~\ref{fig1}a), 
we consider a linear three-atom chain 
with the 1st and the 3rd atoms
placed above the Co (001) surfaces at the ideal positions of
the fcc structure with the distance to the middle-atom of the chain being
$a_0/\sqrt{2}$ which is the nearest neighbor distance in the fcc lattice.
In the second (zigzag-like) configuration (Fig.~\ref{fig1}b), the atomic chain 
is a continuation of the fcc structure along 
[001] direction, thus the symmetry is lower than 
in case of a linear chain. These two configurations 
were chosen to investigate the effect of the
geometrical arrangement on the conductance.

The electronic structure was calculated using the 
non-relativistic spin-polarized version of the screened
Korringa-Kohn-Rostoker (KKR) Green's function method
(for details, see Ref.~\onlinecite{SKKR}).
The potentials were assumed to be spherically symmetric
around each atom. However, the multipole expansion of the charge 
density was taken into account. 
The angular momentum cut-off for the wavefunctions and
the Green's function was chosen to be $l_{\rm{max}}=3$ 
to ensure well converged results.
Within the KKR method\cite{SKKR}, the Green's function of the systems
is obtained in two steps. First, we calculate the Green's function
of the auxiliary system consisting of semi-infinite 
leads separated by a vacuum barrier.
Next, the impurity problem is solved self-consistently
by embedding the chain, surrounding atoms and empty sites 
into the host system, in order to account
for the charge and spin density redistribution effects.
We assumed parallel (P) and antiparallel (AP) magnetic configuration for the Co leads.
The electronic structure of the constriction was calculated
self-consistently for both cases.

Ballistic conductance of the nanocontacts, $g$, was calculated using the Landauer
theory as formulated by Baranger and Stone~\cite{BarStone}.
We consider two semi-infinite Co (001) bulk leads connected 
by the scattering region. 
Conductance is calculated between two atomic planes located 
in the ideal leads.
We neglect tunneling current far away from the contact region
and sum up current contributions 
in real space in the vicinity of the constriction.
Convergence was checked in order 
to achieve errors less than 5\% in the conductance evaluation. 
A detailed description and convergence properties of the method can be found in
Ref.~\onlinecite{conducttheory}. The real space version of the method
employed here was presented also in Ref.~\onlinecite{Papanikolaou}.

In Table I we present magnetic moments calculated for different atomic chains.
We found that magnetic moment is enhanced for the Co
atoms in the Co constriction reaching values up to 2.06~$\mu_B$ for the central
Co atom of the linear chain and 2.28~$\mu_B$ for the zigzag geometry. In the
AP configuration, due to symmetry, the central atoms of the chains
have zero moments and the magnetic profiles are antisymmetric.
In case of Cu, Si, and Al the induced magnetic moments are small as can
be seen from Table~\ref{table1}.                                                             

Our results for the MR at the Fermi energy ($E_F$)
are summarized in Table \ref{table2}. We define a magnetoresistance ratio
as $\mathrm{MR} = (g_P - g_{AP})/g_{AP}$ where $g_P$, $g_{AP}$ are the
conductances for parallel and antiparallel magnetic
configuration of the Co leads.
The calculations predict a MR ratio of 30-40\% for the Co constriction,
and of 20\% in case of the Cu chain suspended 
between the leads. The results for Al and Si chains  
are particularly interesting.
We obtain MR around 50\% for the linear configuration
although magnetic moments are rather small (see Table~\ref{table1}).
The values decrease to 20\% when the symmetry of the constriction is reduced.

\begin{table}[b]
\caption{
Magnetoresistance ratio $MR = (g_P - g_{AP})/g_{AP} \times 100\%$ 
for different zigzag and linear chains suspended between the Co leads.}
\begin{ruledtabular}
\begin{tabular}{ccccc}
{}  & Co & Cu & Al & Si \\ \hline
linear & 29 & 16 & 49 & 50 \\ 
zigzag & 38 & 18 & 19 & 21 
\end{tabular}
\end{ruledtabular}
\label{table2}
\end{table}

\begin{figure}[t]
\begin{center}
\includegraphics[scale = 1.15]{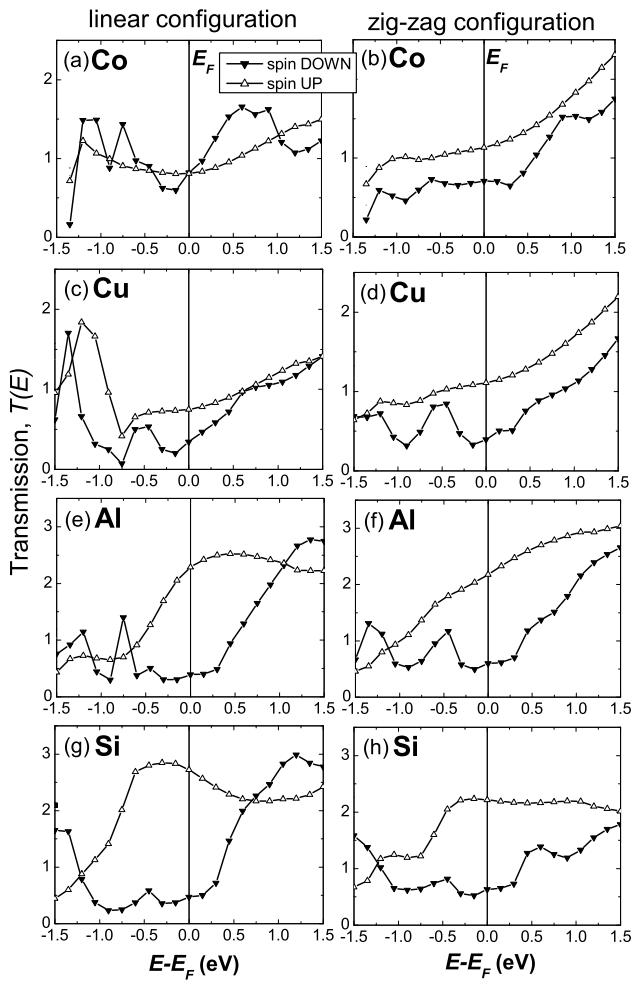}
\caption{\label{fig2}Spin-dependent transmission as a function 
of energy for different linear (left column) and zigzag-like (right column) 
chains suspended between Co electrodes 
with magnetic moments oriented in parallel.
The solid lines with up- and down-triangles correspond to 
transmission for spin-up and spin-down electrons, respectively. }
\end{center}
\end{figure}

\begin{figure}[t]
\begin{center}
\includegraphics[scale = 1.15]{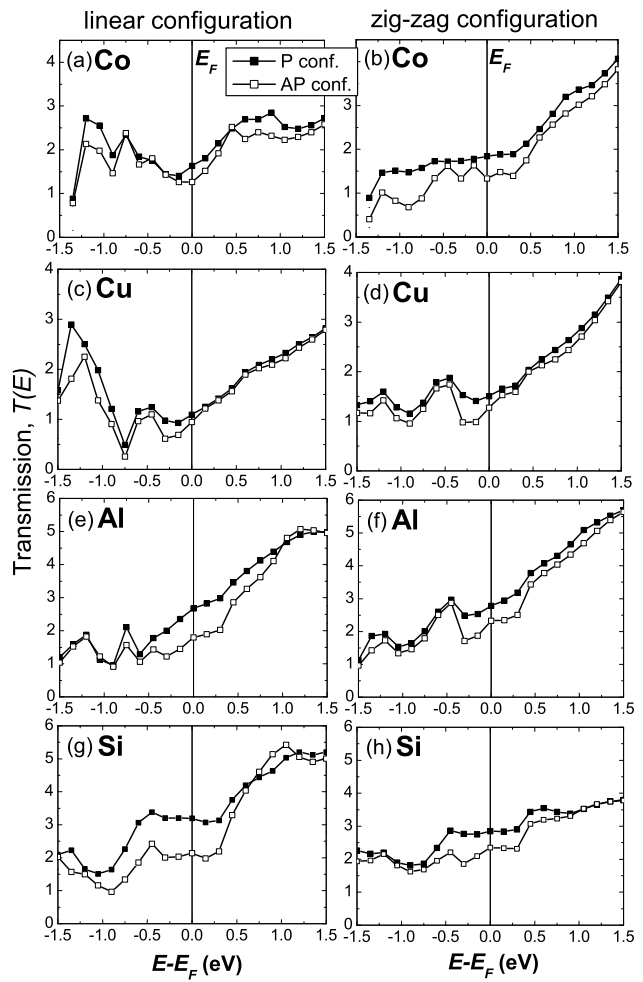}
\caption{\label{fig3}
Total transmission for both spins
as a function of energy for different atomic chains
suspended between the Co leads
in case of parallel (black squares) and antiparallel (open squares)
orientation of magnetic moments of Co. Transmission
for the parallel configuration is the sum of the transmission for spin-up
and spin-down electrons shown in Fig.~\ref{fig2}. In case of the antiparallel 
configuration transmission is the same for both spins.
}
\end{center}
\end{figure}

In order to understand the relation between electronic structure and
transport properties we consider the energy-dependent transmission, $T(E)$,
through the constriction (Fig.~\ref{fig2} and Fig.~\ref{fig3}). 
In the linear response the conductance per spin channel 
is related to the total transmission as 
$g = e^2/h \int_{-\infty}^{+\infty}\{-f'(E)\}T(E)dE$,
where $f'(E)$ is the derivative of the Fermi-Dirac distribution function.
For zero temperature the conductance is $g = e^2/h\; T(E_F)$. 
However, in case of an applied voltage, $V$,
states in the energy window of $eV$ close $E_F$ are relevant for the 
electron transport.

In Fig.~\ref{fig2} we present the transmission of Co, Cu, Al and Si constrictions
between Co leads for both spin channels for parallel magnetic configuration.
We see that transmission for majority (spin-up) electrons is generally
a rather smooth function of energy. However, for minority (spin-down) electrons
the transmission exhibits rather  complicated behavior as a function of energy
caused by Co $d$ states which also contribute to electron transport.
In Fig.~\ref{fig3} we present the total transmission (sum of both spins), for parallel (P) configuration
in comparison with the antiparallel (AP) configuration. 
The conductance of the considered systems is the transmission
at the Fermi energy, $T(E_F)$, in units of $e^2/h$. The difference
of the transmission between P and AP configuration at $E_F$ is a measure
of magnetoresistance.

\begin{figure}[t]
\begin{center}
\includegraphics[scale = 1.0]{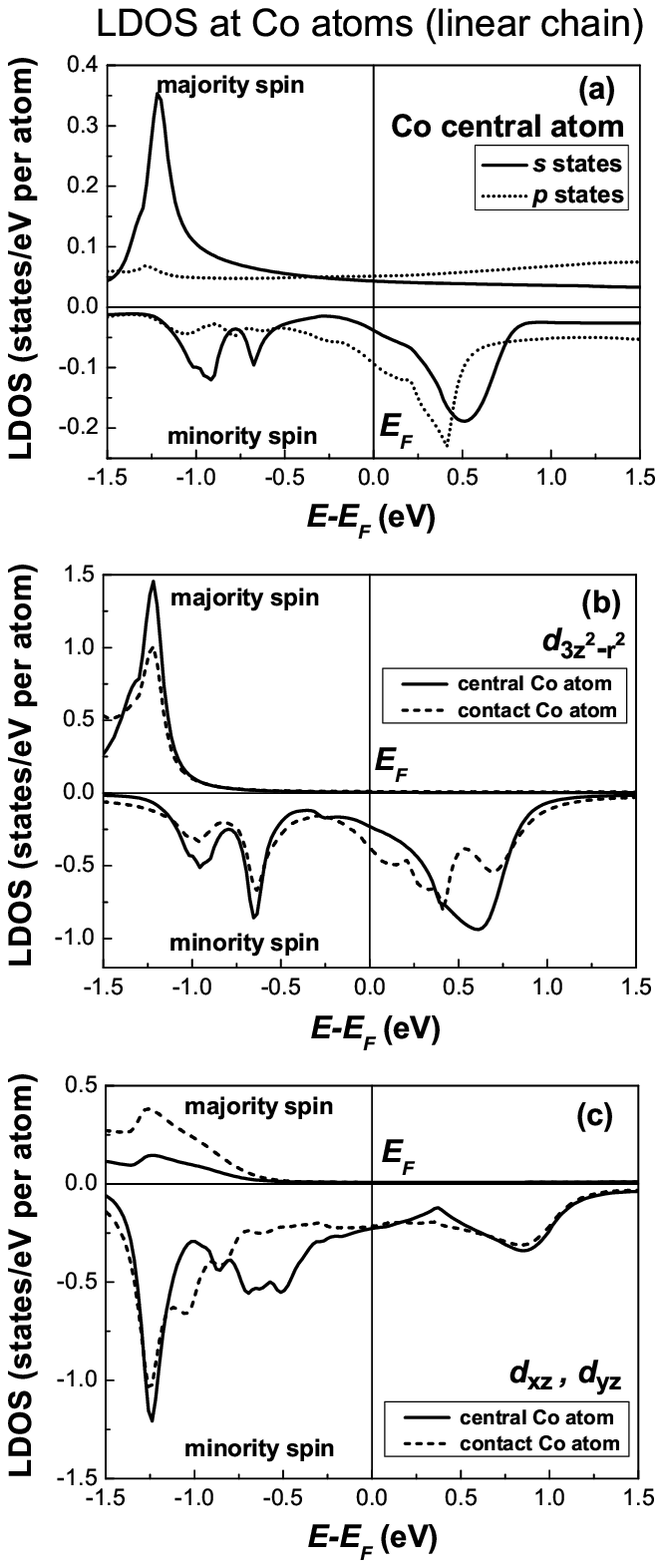}
\caption{
\label{fig4}Symmetry projected local density of states (LDOS)
at the Co atoms of the linear three-atom chain 
for the case of parallel orientation of magnetizations
in the Co leads. We present the $s$, $p$ and all $d$ contributions relevant
for a transport along the wire axis ($z$-axis). 
The contact Co atom is the chain atom next to the surface of the Co lead.
Panel (a): $s$ and $p$ states of the central Co atom; 
panel (b): $d_{3r^2 - z^2}$ states of the central and contact Co atoms;
panel (c): $d_{xz}$ and $d_{yz}$ states of the central and contact Co atoms. } 
\end{center}
\end{figure}

Let us first concentrate on Co constrictions (Fig.~\ref{fig2}a,b).
The energy dependence of the transmission can be interpreted in terms
of local densities of states (LDOS). For this reason,
in Fig.~\ref{fig4} we present the symmetry projected
LDOS, $s$, $p$, $d_{3z^2-r^2}$, $d_{xz}$, $d_{yz}$,
at the Co central atom of the linear chain. In case of linear configuration
the $d_{xy}$, $d_{x^2-y^2}$ orbitals do not support current since
they are oriented perpendicular to the current direction ($z$-axis)
and form very sharp resonances in the LDOS of the central Co site 
of the chain because of weak coupling with 
the orbitals of the neighboring sites.
From Fig.~\ref{fig4}a we see that Co majority states 
close to $E_F$ are mainly $sp$-like since the $d$ band is fully occupied 
and located below $-0.75$~eV with respect to the Fermi energy, 
while for minority states the Fermi level crosses the $d$ band. 
This is also valid for the case of a Co zigzag chain.

The examination of the LDOS at the Co sites of
the linear chain shows that due to the localized and 
directional character of the $d$ orbitals in real space, 
transmission is high only if the orbitals of 
the same symmetry at neighboring sites of the chain are coupled. 
This can be seen from Fig.~\ref{fig4}b and c. For example, the minority
$d_{xz}$ and $d_{yz}$ states at $-1.2$~eV (Fig.~\ref{fig4}c) survive
also at the neighboring atom (dotted line) and causes high transmission.
Similarly, the jump in the transmission around 
$-0.75$~eV for spin-down electrons (Fig.~\ref{fig2}a) correlate with the peak
of the minority $d_{3z^2-r^2}$ state (Fig.~\ref{fig4}b).  

Comparing Figs.~\ref{fig2}a,b we can conclude that 
for Co constrictions the transmission of spin-up electrons 
is rather insensitive to structural changes
simulated by the linear and zigzag configurations of the atomic chains. 
On the contrary, the geometrical arrangement 
of the constriction seems to be more important for  
$d$ electrons. 
Reducing the symmetry of the atomic chain by considering zigzag geometry destroys
coupling between $d$ orbitals, therefore the transmission
of spin-down electrons is reduced and 
we obtain a smooth energy dependence (see Fig.~\ref{fig2}b).                                        

Our study is restricted to collinear magnetic configurations.  
To estimate the influence of this approximation
we have considered also two-atom Co chains where a $180^0$ domain
wall is formed in the antiparallel configuration. 
In that case MR at $E_F$ was found to be
approximately 15\% and the overall behavior is not far from the one
discussed above. Thus, our calculations predict a MR ratio rather
sensitive to the geometrical and magnetic structure. 
In addition, the spherical potential approximation might lead to small shifts 
of the electronic states to different energies, compared to a more accurate 
full-potential description. This might influence the MR values at the Fermi 
level reported in the paper. In particular, in cases where there is a strong 
variation of the transmission with energy close to the Fermi level, the reported 
MR values are expected to be less accurate.
However, we believe
that the configurations under consideration
give a representative order of magnitude of the MR effect 
in Co nanocontacts, but cannot explain the MR ratio of Co contacts
which was recently observed experimentally \cite{Garcia}. 
Moreover, our calculations for Ni nanocontacts using a zigzag configuration (Fig.~\ref{fig1}b)
give a magnetoresistance value of 24\%
contrary to the MR ratios of hundred percent observed in electrodeposited Ni
nanocontacts.  \cite{Garcia}  
Thus, our calculations show that within the accuracy 
of density functional theory the experimental values cannot be explained
assuming a clean, abrupt domain wall and ballistic transport.
Further investigations in particular on the role of defects on magnetotransport
are required, since a recent study indicates a significant enhancement
of the MR ratio if oxygen is assumed to be at the contact region \cite{ninano03}.

\begin{figure}[th]
\begin{center}
\includegraphics[scale = 1.0]{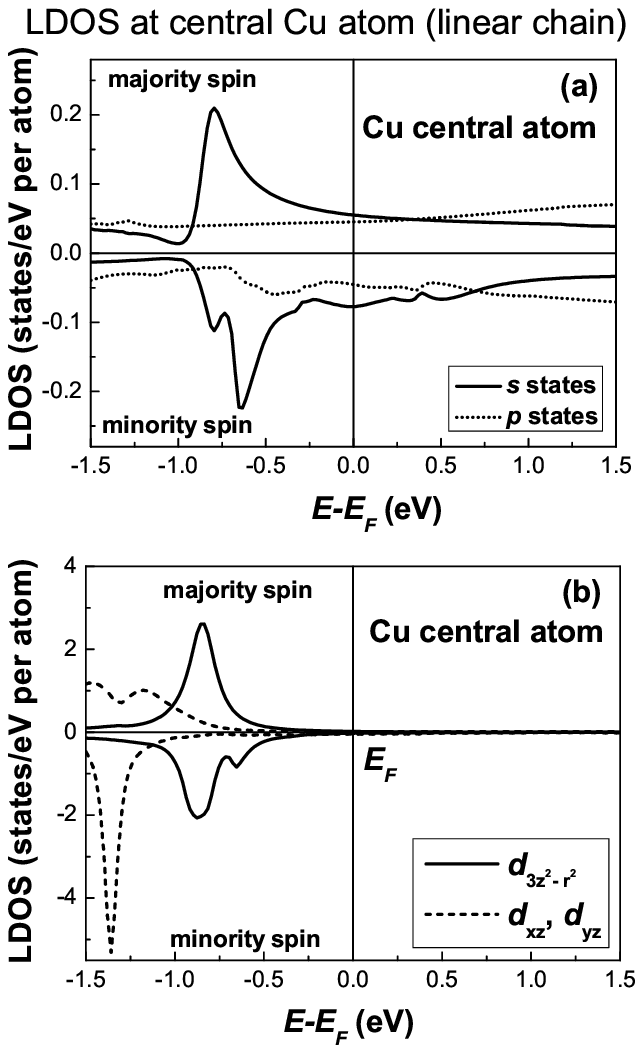}
\caption{
\label{fig5}Symmetry projected local density of states (LDOS)
at the central Cu atom of the linear three-atom chain 
for the case of parallel orientation of magnetizations
in the Co leads. Panel (a): $s$ and $p$ states; 
panel (b): $d_{3r^2 - z^2}$, $d_{xz}$ and $d_{yz}$ states. }
\end{center}
\end{figure}

The next question is whether the electronic structure of the leads 
affects the transport properties through constrictions? 
To answer this question the Co chains were replaced by Cu chains.
It is well known that the majority band of Co matches 
the Cu band whereas the minority band of Co
differs strongly from the Cu band. Our results show that induced magnetic moments
in the Cu chain are quite small (see Table~\ref{table1}) and the
spin-splitting of the LDOS is not significant (see Fig.~\ref{fig5}).   
Consequently, the calculated difference of the transmission for the two 
spins is mainly caused by the Co leads (see Fig.~\ref{fig2}c,d).
In case of the linear configuration 
there are highly conducting states around 
$-1.2$~eV both for spin-up and spin-down channels. 
This is due to the matching of the Co
$d$ band to the Cu $d$ states which are located in the energy
region below $-0.75$~eV (Fig.~\ref{fig5}b).
However, these highly conducting channels are closed in the zigzag configuration
(Fig.~\ref{fig2}d) due to the same symmetry reasons discussed in case of the Co chains. 
Above $-0.75$~eV the electronic structure of
spin-up Co states and Cu states is similar, thus the majority (spin-up) 
transmission of the Cu chains (Fig.~\ref{fig2},c,d) is similar 
to the majority transmission of the Co chains (Fig.~\ref{fig2},a,b).
The energy-dependent transmission for spin-up electrons
is smooth in both configurations. However, the transmission of the spin-down
electrons (Fig.~\ref{fig2}c,d) is significantly reduced
and, finally, a MR ratio of $\sim 15\%$ at $E_F$ 
is obtained (see Fig.~\ref{fig3}c,d). 
The value 1.1~$e^2/h$ of the majority (spin-up) conductance
at~$E_F$ for three-atom zigzag chain of Cu is in agreement
with the conductance of an infinite monoatomic zigzag chain of fcc Cu
oriented along [001] direction.
Since in that case only one band is crossing the Fermi level, 
the conductance is $e^2/h$ per spin channel.  
Finally, we conclude that the electronic structure of the leads
strongly influences the transmission through the constriction. 
The $s$-like states in the leads (majority states of Co) 
show a high transmission through an $s$-like chain, however, 
due to the band mismatch $d$-like minority Co states are strongly
scattered and show a low transmission through the Cu chain.  

\begin{figure}[t]
\begin{center}
\includegraphics[scale = 1.0]{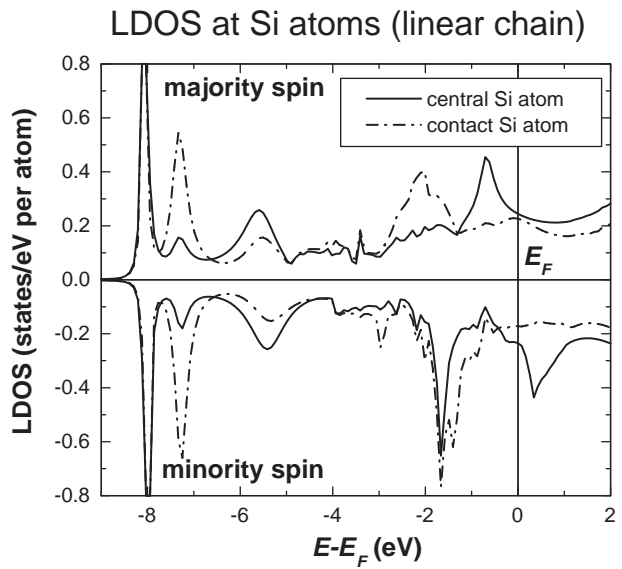}
\caption{
\label{fig6}Symmetry projected local density of states (LDOS)
at the Si atoms of the linear Si chain in case of parallel
orientation of magnetizations in the Co leads. Solid line: central Si atom;
dashed-dotted line: contact Si atom next to the Co surface.  
The peaks of the LDOS around $-1.8$~eV and near $E_F$ have $p_x,p_y$ character.}
\end{center} 
\end{figure}

We further consider
conductance through chains made from $sp$-elements connected to the leads. 
We have chosen Si and Al, 
since Si clusters of 2--20 atoms can be produced in the 
gas phase \cite{Siclusters}, thus measurements similar
to Ref.\onlinecite{vanRuitenbeek_Nature} might be possible. 
Although the induced magnetic moments of Al and Si atoms are small,
the MR ratio can be rather large 
(see Tables~\ref{table1},\ref{table2}). 
This is due to the "splitting" of the
energy-dependent transmission curves 
which reflects the spin-splitting of 
the Co bands (see Figs.~\mbox{Fig.~\ref{fig2}{e--h}}). 
For example, in case of a linear Si chain (Fig.~\ref{fig2}g),
the transmission of spin-up electrons increases up to 3 around $-0.6$~eV 
while the spin-down channel has low transmission 
which rises only above~$E_F$. 
As a consequence, the spin-polarization of the current at $E_F$ is high,  
$P = (g_{\uparrow} - g_{\downarrow}) 
/(g_{\uparrow} + g_{\downarrow})  = 71\%$, 
which causes 50\% MR (Fig.~\ref{fig3}g,h).
The characteristic jump of $\sim2$ in the transmission
that is obtained in both spin channels at different energies
can be explained by means of the spin-resolved LDOS
at the Si sites (Fig.~\ref{fig6}). The majority LDOS of the 
central atom shows a pronounced peak at $-0.6$~eV 
below $E_F$ shaped like a van-Hove singularity in one
dimensional systems. This peak has $p_x,p_y$ character. 
The same $p_x,p_y$ peak is seen in the minority band
but it is shifted due to induced spin-splitting to higher energies 
just above $E_F$. This state is responsible for the increase 
of the transmission by $\sim2$.
The minority LDOS of the central Si atom shows another
pronounced resonance at $-1.8$~eV below $E_F$ which does not occur 
in the majority band and stems from the Co-Si interaction.
This peak is also visible in the minority LDOS of the Si close to the Co
surface. Since in the minority band the Si van-Hove-like state
becomes unoccupied due to spin-splitting, the electrons occupy
the state formed by the Co-Si interaction which leads to nearly
zero magnetic moments in the Si chain.
The same effect of "splitting" of the energy-dependent transmission curves 
is also observed for Al, but the one-dimensional character of
the $p_x, p_y$ state is less pronounced compared to Si, therefore the
increase of the transmission of the linear Al 
chain is smoother (Fig.~\ref{fig2}e). 

In the zigzag configuration the $p_x, p_y$ degeneracy is lifted,
so that for Si (Fig.~\ref{fig2}h) 
the transmission increases in steps of $\sim1$ 
both for spin-up and spin-down channels. 
In case of Al the steps are not so pronounced (Fig.~\ref{fig2}f).
comparing the zigzag with the linear chains we see that the "splitting"
of the transmission curves is conserved in all cases 
(\mbox{Fig.~\ref{fig2}{e--h}}). 
The MR ratio, however, drops for both Al and Si zigzag chains 
(Table~\ref{table2}).

In conclusion, we have presented {\it ab-initio} calculations
of the conductance of magnetic atomic-sized contacts.  
MR ratios up to 50\% are predicted for different three-atom chains 
suspended between Co leads.  Our results show that the conductance 
through $s$-like states is robust against structural changes whereas 
the transmission is strongly influenced by the atomic arrangement in 
the chain if $p$ or $d$ states contribute to the current.  
Furthermore, we have demonstrated that the spin-polarization of the
ferromagnetic leads induces a spin-polarization of the current although
the induced magnetic moments in the chain could be rather small (Cu, Al, Si).   
The induced spin-splitting of the wire states due to ferromagnetic leads
is practically zero for $s$~states but could be as large as 1~eV for $p$~states. 
In case of $sp$ chains (Al, Si) this causes strong spin-polarization of the 
current and, consequently, large MR values.

\enlargethispage{\baselineskip}

\end{document}